\documentclass[aps,prd,twocolumn,preprintnumbers,nofootinbib,superscriptaddress,amsmath]{revtex4-2}
\usepackage{subcaption} 
\usepackage{graphicx}
\usepackage{amsmath}
\usepackage{amssymb}
\usepackage{amsfonts}
\usepackage{hyperref}
\usepackage{url}
\usepackage{color}
\usepackage{bm}
\usepackage{array}
\usepackage{ragged2e}
\usepackage{comment}

\newcommand{\nv}{\boldsymbol{\hat{n}}}

\begin{document}

\preprint{APS/123-QED}

\title{Skew-spectra: a generalization to spin-$s$}
\author{
  Alexander Roskill
}
 \email{alexander.roskill@physics.ox.ac.uk}\affiliation{
  Astrophysics, University of Oxford, Denys Wilkinson Building, Keble Road, Oxford OX1 3RH, United Kingdom
}
\author{
  Sara Maleubre 
} 
\affiliation{
 Astrophysics, University of Oxford, Denys Wilkinson Building, Keble Road, Oxford OX1 3RH, United Kingdom
}
\author{
  David Alonso 
} 
\affiliation{
Astrophysics, University of Oxford, Denys Wilkinson Building, Keble Road, Oxford OX1 3RH, United Kingdom
}
\author{
  Pedro G. Ferreira
} 
\affiliation{
  Astrophysics, University of Oxford, Denys Wilkinson Building, Keble Road, Oxford OX1 3RH, United Kingdom
}

\date{\today}

\begin{abstract}
Skew-spectra allow us to extract non-Gaussian information by taking the square of a map and finding the power spectrum of this new map with the original map. This allows us to use much of the infrastructure of power spectra and avoid the intricacies of estimating three point statistics. In this paper we present the first extension of skew-spectra to arbitrary spin-$s$ fields, as a means to extract non-Gaussian information efficiently from cosmological data sets like cosmic shear or CMB polarization. We apply the formalism to weak lensing in the context of large scale structure, and discuss different ways of combining fields to build skew-spectra, all while avoiding the problems associated with mass-mapping. We provide plots of these new statistics for $\Lambda$CDM and vary cosmological parameters. 

\end{abstract}

\maketitle

\section{Introduction\label{sec:intro}}
The analysis of cosmological data sets is a daunting task of considerable technical challenge and computational expense \cite{Tegmark_1997}. This is ringing true in the current era of so-called precision cosmology, where the field has transitioned from lacking in data to having a wealth of data, of unprecedented depth and resolution, coming from a host of surveys such as DESI \cite{DESICollaboration_2022,desicollaboration2025desidr2resultsii}, LSST \cite{Ivezić_2019}, Euclid \cite{laureijs2011eucliddefinitionstudyreport}, Nancy Grace Roman Space Telescope \cite{spergel2013wfirst24astronomerknow,foley2018lsstobservingstrategywhite}, and SPHEREx \cite{Crill_2020}. The standard approach, which also helps to combat the challenges introduced by this shift in data volume, is to compress the data into `summary statistics'. These are then used in parameter inference to constrain cosmology \cite{Ferreira:2025hwc}. Many surveys focus their efforts on the two-point correlation function or its Fourier counterpart, the power spectrum, which completely describes the statistics of a Gaussian field. This fact makes the power spectrum a highly valuable tool for studying the cosmic microwave background (CMB), since the early universe was very smooth \cite{planckcollaboration2019planck2018resultsix,Liguori_2010,Yadav_2010,Chen_2010}. However, as the universe evolved under non-linear gravitational forces, the matter in the universe began to collapse into the structure that we see today. This highly non-Gaussian structure contains huge amounts of information regarding the fundamental physics that drove it to its current topology.

There are a plethora of (higher-order) statistics which are constructed to extract non-Gaussian information from data sets. In practice, the statistics we choose to employ should satisfy certain considerations. Firstly, we require the statistics to be informative. This means that measuring them will provide tight constraints on cosmological parameters of interest and should ideally only be limited by the fundamental noise in the data, nuisance parameters, and systematics. Many higher-order statistics are motivated by theoretical and phenomenological arguments, with no guarantee to optimally extract the information. Examples include the use of cosmic void statistics, which are able to probe the nature of dark energy and neutrinos \cite{pisani2019cosmicvoidsnovelprobe} or the four-point correlation function, which is sensitive to parity violating physics \cite{cahn2021testcosmologicalparityviolation,hou_measurement_2023,Cabass_2023,coulton2023signaturesparityviolatinguniverse,Philcox_2022,Cabass_2023}. Other statistics are designed to capture the full non-Gaussian information,  such as the $k$-Nearest Neighbor Cumulative Distribution Functions \cite{Banerjee_2020}, and convolutional neural networks (CNNs). However, many of these lack interpretability and are difficult to apply to real data, so work has been done to bridge the gap between these and traditional statistics using for example wavelet scattering transforms \cite{6522407,mallat2012groupinvariantscattering,Valogiannis_2022}. Despite potentially extracting new information from data sets, complex statistics present difficulties in theoretical modeling, particularly in statistics where perturbation theory \cite{Bernardeau_2002} cannot be applied or in regimes where it is no longer valid \cite{COORAY_2002,rimes_information_2005,Neyrinck_2006, 2016PhLB..762..247N}. In such cases, more sophisticated approaches, including emulators \cite{Sui_2025,Angulo:2020vky,2021_euclid_emu}, fitting functions \cite{Smith_2003,Takahashi_2020}, and simulation-based inference (SBI) \cite{marin2011approximatebayesiancomputationalmethods}, are often employed. These tasks become considerably more challenging as the statistical measures increase in complexity.

The second characteristic that we would like the statistics to have is that we should be able to implement them in highly efficient ways, which is of particular importance given the volume of data \cite{Heavens_2017}. Even optimally calculating power spectra of real data sets becomes impractical except on relatively large scales, and methods such as the Pseudo-$C_\ell$ \cite{Hivon_2002,Alonso_2019} and FKP estimators \cite{1994ApJ...426...23F}, which trade optimality for efficiency\footnote{ For example, the Pseudo-$C_\ell$ measurement complexity scales as $\mathcal{O}(\ell_{\rm{max}}^3)$ in contrast to Quadratic Maximimum Likelihood (QML) estimators \cite{Tegmark:2001zv} which scales as $\mathcal{O}(\ell_{\rm{max}}^6)$ (albeit QML estimators optimally extract the two-point information of a given field).}, are often employed. The size of the data vector under consideration is also critical in this regard. For example, in traditional analysis pipelines, which use Markov Chain Monte Carlo (MCMC) techniques, a covariance of the data vector is required. This often necessitates an abundance of costly $N$-body simulations, which means that measurement of the statistic from data itself should be efficient. Even modern analysis techniques such as SBI \cite{ Alsing_2019, Cranmer_2020}, in which machine learning is drawn upon to learn the likelihood, similarly rely on large numbers of simulations \cite{Alsing_2018} to train neural density estimators \cite{2019arXiv190604032D}. This problem is worsened by introducing more complex data vectors. Much effort has gone into developing compression techniques to reduce the computational load in an attempt to begin to use higher-order statistics (HOS) on real data \cite{Zablocki_2016,Akhmetzhanova_2023,PhysRevD.97.083004}. Most notable of which is the data compression algorithm MOPED \cite{Heavens_2000,Heavens_2017}.  

A further consideration is ensuring that statistical analyses remain robust to the impact of complex survey geometry and observational systematics. With forthcoming surveys set to explore ever more distant regions of the Universe at unprecedented resolution and low noise, ever tighter control of these systematics will be crucial, as they are likely to become the dominant limiting factor. This is particularly important in photometric surveys, which are generally deep and affected by contamination from dust extinction, stars, and a plethora of observing conditions \cite{2002ApJ...579...48S,2011MNRAS.417.1350R,2013MNRAS.432.2945H}. Other problems arise in weak lensing HOS, where many statistics (eg. peak counts \cite{marian2009cosmology,Dietrich:2009jq,2024MNRAS.534.3305H,Marques_2024,Grand_n_2024}, probability density functions \cite{Liu_2019,Barthelemy_2020,Boyle_2021,PhysRevD.102.123545,PhysRevD.108.123526,Castiblanco_2024}, $\ell_1$-norm \cite{Ajani_2021,2024A&A...691A..80S}, Minkowski functionals \cite{PhysRevD.85.103513,PhysRevD.91.103511,2020A&A...633A..71P,Armijo_2025} and Betti numbers \cite{Parroni_2021}) rely on mass-mapping, where the observed (reduced) shear of galaxies are translated into convergence maps. Methods such as the Kaiser-Squires algorithm \cite{kaiser1993mapping} are known to produce spurious signals in the inversion due to mask effects \cite{Shirasaki:2013eua,liu2014mask}, thereby necessitating inpainting methods \cite{Starck:2021wnm,Marques_2024}. The choice of mass-mapping can also have large effects on the final parameter constraints \cite{refId0}.

One of the simplest higher-order statistics is the bispectrum, the Fourier-space analog of the three-point correlation function, with robust results made possible by theoretical predictions that link them to fundamental quantities such as the matter overdensity \cite{Bernardeau_2002}. Furthermore, the large scale structure (LSS) bispectrum may be used to improve bounds on primordial non-Gaussianity \cite{PhysRevLett.86.1434,Chen_2021}. In galaxy surveys the galaxy bispectrum has been used to break degeneracies between galaxy bias and the amplitude of dark matter fluctuations \cite{PhysRevLett.86.1434,1993ApJ...413..447F,Matarrese_1997,Scoccimarro_2001,Verde_2002,Smith_2006,Gilmar_bias,Nishimichi_2007,Sugiyama_2018,Sugiyama_2020,Philcox_2022,harscouet2025constraintscmblensingtomography,verdiani2025cosmologicalconstraintsangularpower}. Furthermore, it has helped tighten constraints on cosmology and measure independently the baryon acoustic oscillations (BAO) features in the LSS \cite{Slepian_2017}. There has also been much work concerned with jointly constraining $\Omega_m$ and $\sigma_8$ \cite{Gil_Mar_n_2015,Sefusatti_2006,harscouet2025constraintscmblensingtomography,verdiani2025cosmologicalconstraintsangularpower}, including through the use of the weak lensing bispectrum \cite{Hui_1999,Cooray_2001,Takada_2004,Coulton_2019,Munshi_2020}, which probes the non-linear regime in a way which is free from the issues associated with galaxy bias \cite{COORAY_2002}.  Other examples include testing the equivalence principle \cite{Umeh_2021}, parity violation \cite{PhysRevD.83.027301}, and constraining the sum of neutrino masses \cite{Hahn_2020}. 

Despite its successes, there are large costs associated with trying to use the bispectrum in realistic data pipelines with control over systematics. There have been many approaches to deal with this, and of note are the proxy statistics for the bispectrum. These are statistics which are simpler to interpret and compute, but still contain bispectrum information. Many studies focus for example on the squeezed configurations (and related `integrated bispectra') which contain a large fraction of the non-Gaussian information of the field \cite{Chiang_2014,Barreira_2019,Biagetti_2022,salvalaggio2024bispectrumnongaussiancovarianceredshift,giri2023constrainingfnlusinglargescale}. A comparison of these can be found in Ref.~\cite{Byun_2017}. Another often used proxy bispectrum estimator is the skew-spectrum, constructed by taking the square of a given field and estimating its cross-power spectrum with the original field. This has been applied to the 3D bispectrum \cite{Schmittfull_2015,Dizgah_2020,Munshi_2022,Chakraborty_2022}, but also to the angular bispectrum of the CMB in Ref.~\cite{Cooray_2001} and to the galaxy-galaxy-CMB lensing bispectrum in Ref.~\cite{farren2023detectioncmblensing,harscouet2025constraintscmblensingtomography}. Recently, a new skew-spectrum for scalar fields involving an additional filtering step has been developed in Ref.~\cite{Harscouet2025Fast} and applied to the cross-correlation of CMB lensing and galaxy clustering in Refs.~\cite{harscouet2025constraintscmblensingtomography} and \cite{verdiani2025cosmologicalconstraintsangularpower}. These statistics have a key advantage: the large infrastructure built around the efficient estimation of power spectra and their uncertainties in the presence of complex survey geometry and systematics can be brought to bear to achieve a fast, accurate, and reliable estimator of the bispectrum and its covariance matrix \cite{Harscouet2025Fast}.

The goal of this paper is to explore what we can do to extend skew-spectra to spin-$s$ fields. We describe the mathematics of spin-$s$ fields in Sec.~\ref{sec:general_formalism}. We develop the skew-spectra statistics in a general form, considering arbitrary cross correlations of spin-$s$ fields in Sec.\ref{sec:spin-squared_maps} (enabling its potential application to any cosmological problem involving spin-$s$ fields). We discuss how these statistics are formed by various linear combinations of bispectrum types. We then apply our formalism to weak lensing in Sec.~\ref{sec:weak_lensing_spin_fields}, describing its relation to the angular bispectrum, as well as the cosmology dependence. We highlight that one of the main advantages of this process is that we work directly with the shear field, in contrast to many of the weak lensing HOS, which rely on mass-mapping methods. In addition, our formalism leans on the highly-developed infrastructure of power spectra and in particular the Pseudo-$C_\ell$ \cite{Alonso_2019} approach of correcting for the effects of masks and systematics, thereby incorporating many of the important features of HOS (informative, efficient to measure and predict, robust to masks, etc). We conclude in Sec.~\ref{sec:discussion}. \\

\section{Spin-s fields\label{sec:general_formalism}}
In cosmology we are often concerned with tensor fields (eg. weak lensing or polarization in the CMB) projected onto the celestial sphere. These fields are typically assumed to be isotropic and therefore invariant under two-dimensional rotations, i.e. $SO(2)$. As such it is physically insightful to decompose these tensor objects into their irreducible representations under $SO(2)$, and study these individual components. Due to the isomorphism between $SO(2)$ and $U(1)$ it is simpler to work with complex fields, rather than two dimensional real fields. This leads us to define the concept of a spin-$s$ field \cite{10.1063/1.1931221,10.1063/1.1705135}.

A spin-$s$ field, $a$, with spin $s_a$, is a complex field defined through its transformation under rotations around $\nv$, as 
\begin{equation}
    a(\nv) \rightarrow a(\nv) \  e^{is_a\psi},
\end{equation}
where the rotation is by the angle $\psi$. The harmonic decomposition is 
\begin{equation}\label{eq:harmonic_decomp_spin}
\begin{split}
    a (\nv) &= \sum_{\ell m} \ {}_{s_a}a_{\ell m} 
 \ {}_{s_a}Y_{\ell m}(\nv),\\
 {}_{s_a}a_{\ell m}  &= \int \mathrm{d}\nv \ a(\nv) \ {}_{s_a}Y^{*}_{\ell m}(\nv),
\end{split}
\end{equation}
where ${}_{s}Y_{\ell m}\equiv\sqrt{(\ell-s)!/(\ell+s)!}\eth^sY_{\ell m}$ are the spin-weighted spherical harmonics (SWSH), $Y_{\ell m}$ are the standard spherical harmonics and
\begin{equation}\label{eq:spin_raising}
    \eth{}\ a\equiv -\sin^{s_a}\theta\left(\partial_{\theta}+i\frac{\partial_{\phi}}{\sin\theta}\right)\sin^{-s_a}\theta  \ a,
\end{equation}
\begin{equation}\label{eq:spin_lowering}
    \bar{\eth}\ a\equiv -\sin^{-s_a}\theta\left(\partial_{\theta}-i\frac{\partial_{\phi}}{\sin\theta}\right)\sin^{s_a}\theta  \ a,
\end{equation}
are the spin-raising and spin-lowering operators respectively acting on a spin-$s_a$ field, $a$. As their names suggest, the application of $\eth$ ($\bar{\eth}$) on a given field raises (lowers) its spin by $1$. These are written in the standard orthonormal spherical coordinates. We work instead with the $E$ and $B$ modes \cite{Seljak_1997,Kamionkowski_1997} of the $a$ field 
\begin{equation}\label{eq:e_modes}
    E^{a}_{\ell m} \equiv -\frac{1}{2}\left({}_{s_a}a_{\ell m} + (-1)^{s_a}{}_{-s_a}a_{\ell m}\right),
\end{equation}
\begin{equation}\label{eq:b_modes}
    B^{a}_{\ell m} \equiv -\frac{1}{2i}\left({}_{s_a}a_{\ell m} - (-1)^{s_a}{}_{-s_a}a_{\ell m}\right),
\end{equation}
which form the harmonic coefficients of the scalar and pseudo-scalar quantities
\begin{equation}\label{eq:e_mode_exp}
    E^a(\nv) = \sum_{\ell m} E^{a}_{\ell m} Y_{\ell m}(\nv),
\end{equation}
\begin{equation}\label{eq:b_mode_exp}
    B^a(\nv) = \sum_{\ell m} B^{a}_{\ell m} Y_{\ell m}(\nv).
\end{equation}
They transform as even-parity and odd-parity functions respectively
\begin{equation}
    E^a(-\nv) = E^a(\nv), \ \ \ B^a(-\nv) = -B^a(\nv),
\end{equation}
which implies that the coefficients transform as 
\begin{equation}\label{eq:parity_eandb}
    \mathbb{P}(E^{a}_{\ell m}) = (-1)^\ell E^{a}_{\ell m}, \ \ \mathbb{P}(B^{a}_{\ell m}) = (-1)^{\ell+1} B^{a}_{\ell m}, 
\end{equation}
where $\mathbb{P}$ is the parity operator ($\nv\rightarrow-\nv$). This result may be shown straightforwardly using the parity properties of the spherical harmonics $Y_{\ell m}(-\nv) = (-1)^{\ell}Y_{\ell m}(\nv) $. Due to their construction, by taking an integral over the full sky in Eq.~(\ref{eq:harmonic_decomp_spin}), the $E$- and $B$-mode fields are inherently non-local quantities.

To make the physical degrees of freedom evident, while compressing the information content of fields, we impose statistical isotropy, thereby making the harmonic space correlations invariant under rotations \cite{Hu_2000}. We can consider the power spectrum between two fields, $a$ and $b$, with spins $s_a$ and $s_b$ respectively
\begin{equation}\label{eq:spins_ps}
    \langle {}_{s_a}a_{\ell m }({}_{s_b}b_{\ell'm'})^*  \rangle \equiv \delta_{\ell \ell'} \delta_{mm'}C^{ab}_{\ell}. 
\end{equation}
\begin{equation}
    \langle {}_{s_a}a_{\ell m }({}_{-s_b}b_{\ell'm'})^*  \rangle \equiv \delta_{\ell \ell'} \delta_{mm'}\bar C^{ab}_{\ell}
\end{equation}
Alternatively, we can separate the quantities into power spectra for $E$ and $B$ modes  
\begin{equation}
    \langle F^{a}_{\ell m}(G^{b}_{\ell' m'})^*\rangle\equiv\delta_{\ell\ell'}\delta_{mm'}C^{F^{a}G^{b}}_{\ell},
\end{equation}
where $F$ and $G$ represent either the $E$ or $B$ modes. We use this notation  throughout. We can clearly see, using Eq.~(\ref{eq:parity_eandb}), which statistics are parity even and odd 
\begin{equation}\label{eq:power_parity}
    \mathbb P \left(C_{\ell}^{F^aG^b}\right) = P_{FG} C_{\ell}^{F^aG^b},
\end{equation}
where $P_{EE}= P_{BB}= 1$ and $P_{EB}=P_{BE}=-1$. We can also prove using Eqs.~(\ref{eq:e_modes}) and (\ref{eq:b_modes}), that $E^{(a^*)} = (-1)^{s_a}E^a$ and $B^{(a^*)}=(-1)^{s_a+1}B^a$. This implies that we need not consider the complex conjugate of the fields we use to construct the $E$ and $B$ modes, since the statistics formed from these are proportional to one another. We can relate the two types of power spectra, using Eqs.~(\ref{eq:e_modes}), (\ref{eq:b_modes}), as
\begin{equation}\label{eq:power_spin_to_eb}
    C_{\ell}^{ab} = C^{E^aE^b}_{\ell} +C^{B^a B^b}_{\ell} +i\left(C_{\ell}^{E^bB^a}-C_{\ell}^{E^aB^b}\right), 
\end{equation}
\begin{equation}\label{eq:power_spin_to_eb_complex}
    \bar C_{\ell}^{ab} = (-1)^{s_b}\biggl(C^{E^aE^b}_{\ell}-C^{B^a B^b}_{\ell} +i\left(C_{\ell}^{E^bB^a}+C_{\ell}^{E^aB^b}\right)\biggl).
\end{equation}
We can see that the imaginary part contains the parity-odd contribution and the real part contains the parity-even contribution.

We now turn to three point statistics. Imposing isotropy \cite{Hu_2000,Hu_2001}, the three-point harmonic-space correlation function of three scalar fields can be written as
\begin{equation}\label{eq:full_bispectrum}
    \langle F^{a}_{\ell_1 m_1}G^{b}_{\ell_2 m_2} H^c_{\ell_3 m_3}\rangle = \begin{pmatrix}
        \ell_1 & \ell_2 & \ell_3 \\
        m_1 & m_2 & m_3
    \end{pmatrix}
\mathcal{B}^{F^a G^b H^c}_{\ell_1 \ell_2 \ell_3}.
\end{equation}
Again $F$, $G$ and $H$ represent either the $E$ or $B$ modes. The three-point information of the fields is contained in the angular bispectrum, $\mathcal{B}^{F^a G^b H^c}_{\ell_1 \ell_2 \ell_3}$. The `matrices' in the expression above represent the Wigner-$3j$ symbols, and express the rotation invariance of the statistics and allow us to extract the physical content of the fields under this assumption. They satisfy among other things the triangular condition $|\ell_1-\ell_2|\leq \ell_3 \leq \ell_1+\ell_2$, and $m_1+m_2+m_3=0$. An important property of the Wigner-$3j$ is that if $m_1=m_2=m_3=0$ it is only non-zero if $\ell_1+\ell_2+\ell_3=\text{even}$. Other useful properties of the Wigner-$3j$ symbols are included in Appendix \ref{sec:spin_fields}. For a given bispectrum (eg. $EEE$), there are even and odd parity configurations in $\ell$-space \cite{Shiraishi_2014}. Consider how the bispectrum changes under parity, using Eq.~(\ref{eq:parity_eandb}), as 
\begin{equation}\label{eq:bispectrum_parity}
    \mathbb{P}\left(\mathcal{B}^{F^a G^b H^c}_{\ell_1 \ell_2 \ell_3}\right) = P_{FGH}(-1)^{\ell_1+\ell_2+\ell_3}\mathcal{B}^{F^a G^b H^c}_{\ell_1 \ell_2 \ell_3},
\end{equation}
where $P_{FGH}=1$ for an even number of $B$ modes and $P_{FGH}=-1$ otherwise. For an even number of $B$ modes we have an even parity bispectrum if $\ell_1+\ell_2+\ell_3$ is even and an odd bispectrum if $\ell_1+\ell_2+\ell_3$ is odd. Similarly, for an odd number of $B$ modes we have an even parity bispectrum if $\ell_1+\ell_2+\ell_3$ is odd and an odd bispectrum if $\ell_1+\ell_2+\ell_3$ is even.

In the literature the bispectrum is usually restricted to even $\ell_1+\ell_2+\ell_3$ \cite{PhysRevD.83.027301}. In this case, the bispectrum is better expressed as a `reduced bispectrum', $b^{F^a G^b H^c}_{\ell_1 \ell_2 \ell_3}$, defined as
\begin{equation}\label{eq:reduced_bispectrum}
   \langle F^{a}_{\ell_1 m_1}G^{b}_{\ell_2 m_2} H^c_{\ell_3 m_3}\rangle = \mathcal{G}_{\ell_1\ell_2\ell_3}^{m_1 m_2 m_3}
b^{F^a G^b H^c}_{\ell_1 \ell_2 \ell_3},
\end{equation}
where the Gaunt factor is given by
\begin{equation}\label{eq:Gaunt_factor}
\begin{split}
    \mathcal{G}^{m_1 m_2 m_3}_{\ell_1 \ell_2 \ell_3} &= \sqrt{\frac{(2\ell_{1}+1)(2\ell_{2}+1)(2\ell_{3}+1)}{4\pi}}  \times  \\
    &\times\begin{pmatrix}
                \ell_{1} & \ell_2 & \ell_3\\
                0 & 0 & 0
                \end{pmatrix}\begin{pmatrix}
                \ell_1 & \ell_2 & \ell_3\\
                m_1 & m_2 & m_3
                \end{pmatrix}.
\end{split}
\end{equation}
The first Wigner-$3j$ symbol enforces the condition that $\ell_1+\ell_2+\ell_3=\text{even}$. However, it is well known that the configurations $\ell_1+\ell_2+\ell_3=\text{odd}$ could also be sourced for example by tensor perturbations in the case of the CMB temperature bispectrum \cite{PhysRevD.83.027301}.

\section{Formalism for generalized skew-spectra}\label{sec:spin-squared_maps}
In this section, we assume only homogeneity and isotropy as fundamental principles, and present the mathematical structure that extends the notion of skew-spectra to spin-$s$ fields in a probe-agnostic way. For example, we do not assume $\ell_1+\ell_2+\ell_3=\text{even}$ or negligible $B$-modes. We consider the product of spin-$s$ fields, $a \cdot b $, where $a$ and $b$ are two spin-$s$ fields with spins $s_a$ and $s_b$ respectively. The composed field has spin $s_a+s_b$. We can then calculate the $E$ and $B$ modes of the new field and calculate various power spectra. As is possible with skew-spectra of the density field \cite{Harscouet2025Fast}, we are therefore able to access bispectrum information quickly whilst using the significant machinery that has been developed for calculating power spectra. We explore this idea further for weak lensing in Sec.~\ref{sec:weak_lensing_spin_fields}, using the results derived in this section.

We want to calculate and make theoretical predictions for the cross-power spectrum of a map with the new `spin-squared' maps in terms of the $E$ and $B$ modes 
\begin{equation}\label{eq:cross_power}
    \langle F^{a}_{\ell' m'} (G^{b\cdot c}_{\ell m})^*\rangle  \equiv \delta_{\ell\ell'}\delta_{mm'} C_{\ell}^{F^{a}  G^{b \cdot c}},
\end{equation}
where the labels, $a,b,c$, denote the arbitrary spin-$s$ fields we are considering, with spins $s_a$, $s_b$ and $s_c$ respectively. $F$ and $G$ represent either $E$ or $B$ modes. To this end, we are interested in how these power spectra can be written in terms of the angular bispectrum, $\mathcal{B}_{\ell \ell_1\ell_2}$. We demonstrate how this is done in general for the spin-squared maps. Consider expanding $(a\cdot b)(\nv)$ into SWSHs
\begin{widetext}
\begin{equation}
\begin{split}
    &(a \cdot b)(\nv) = \sum_{\ell_1 m_1}\sum_{\ell_2 m_2} {}_{s_a}a_{\ell_1 m_1} 
 \ {}_{s_b}b_{\ell_2 m_2}  \ {}_{s_a}Y_{\ell_1 m_1} \ {}_{s_b}Y_{\ell_2 m_2} \\
 &=  \sum_{\ell_1 m_1\ell_2 m_2} \sum_{\ell m} {}_{s_a}a_{\ell_1 m_1} 
 \ {}_{s_b}b_{\ell_2 m_2}  (-1)^{m +s_a+s_b} \begin{pmatrix}
     \ell_1 & \ell_2 &\ell \\
     m_1 & m_2 & -m
 \end{pmatrix} \begin{pmatrix}
     \ell_1 & \ell_2 & \ell \\
     -s_a & -s_b & s_a +s_b
 \end{pmatrix}\sqrt{\frac{(2\ell_1+1)(2\ell_2+1)(2\ell+1)}{4\pi  }} \ {}_{s_a+s_b}Y_{\ell m},
\end{split}
\end{equation}
where we used Eq.~(\ref{eq:spin_coupling}) to describe the product of SWSHs in terms of one SWSH. The second Wigner-$3j$ can be seen as enforcing the condition of conserving the spin of the product of the fields ($s_a+s_b$ is the spin of the new field $a\cdot b$). We are then able to directly identify  ${}_{s_a+s_b}(a \cdot b)_{\ell m}$ using Eq.~(\ref{eq:harmonic_decomp_spin}). Repeating the steps taken to expand $(a\cdot b)(\nv)$ for $(a\cdot b)^*(\nv)$, we also find ${}_{-(s_a+s_b)}(a \cdot b)_{\ell m}$. Moreover, we know how ${}_{\pm s_a}a_{\ell_1 m_1}$ are related to their $E$ and $B$ modes through Eqs.~(\ref{eq:e_modes}) and (\ref{eq:b_modes}) and are therefore able to relate the $E$ and $B$ modes of the spin-squared fields to the $E$ and $B$ modes of the original fields used to build the spin-squared maps. We find
\begin{equation}
\begin{split}
    &E^{a \cdot b}_{\ell m} = -\frac{1}{2}\sum_{\ell_1 \ell_2 m_1 m_2} \biggl(\left(E_{\ell_1 m_1}^{a}E_{\ell_2 m_2}^{b} - B_{\ell_1 m_1}^{a}B_{\ell_2 m_2}^{b} +iE_{\ell_1 m_1}^{a}B_{\ell_2 m_2}^{b}  +i B_{\ell_1 m_1}^{a}E_{\ell_2 m_2}^{b}\right)\\
    &+(-1)^{\ell_1+\ell_2+\ell}\left(E_{\ell_1 m_1}^{a}E_{\ell_2 m_2}^{b} - B_{\ell_1 m_1}^{a}B_{\ell_2 m_2}^{b} -iE_{\ell_1 m_1}^{a}B_{\ell_2 m_2}^{b}  -i B_{\ell_1 m_1}^{a}E_{\ell_2 m_2}^{b}\right)
    \biggl) (-1)^{m+s_a+s_b}\times \\
    &\times\begin{pmatrix}
        \ell_1 & \ell_2 & \ell \\
        -s_a & -s_b & s_a+s_b 
    \end{pmatrix}\begin{pmatrix}
        \ell_1 & \ell_2 & \ell \\
        m_1 & m_2 & -m 
    \end{pmatrix}
    \sqrt{\frac{(2\ell_1+1)(2\ell_2+1)(2\ell+1)}{4\pi  }},\\
\end{split}
\end{equation}
\begin{equation}
\begin{split}
    &B^{a \cdot b}_{\ell m} = -\frac{1}{2i}\sum_{\ell_1 \ell_2 m_1 m_2} \biggl(\left(E_{\ell_1 m_1}^{a}E_{\ell_2 m_2}^{b} - B_{\ell_1 m_1}^{a}B_{\ell_2 m_2}^{b} +iE_{\ell_1 m_1}^{a}B_{\ell_2 m_2}^{b}  +i B_{\ell_1 m_1}^{a}E_{\ell_2 m_2}^{b}\right)\\
    &-(-1)^{\ell_1+\ell_2+\ell}\left(E_{\ell_1 m_1}^{a}E_{\ell_2 m_2}^{b} - B_{\ell_1 m_1}^{a}B_{\ell_2 m_2}^{b} -iE_{\ell_1 m_1}^{a}B_{\ell_2 m_2}^{b}  -i B_{\ell_1 m_1}^{a}E_{\ell_2 m_2}^{b}\right)
    \biggl) (-1)^{m+s_a+s_b}\times \\
    &\times\begin{pmatrix}
        \ell_1 & \ell_2 & \ell \\
        -s_a & -s_b & s_a+s_b 
    \end{pmatrix}\begin{pmatrix}
        \ell_1 & \ell_2 & \ell \\
        m_1 & m_2 & -m 
    \end{pmatrix}
    \sqrt{\frac{(2\ell_1+1)(2\ell_2+1)(2\ell+1)}{4\pi  }},\\
\end{split}
\end{equation}
where we have used the time reversal property of the Wigner $3j$ of Eq.~(\ref{eq:time_reversal}). Note that the $B^{a\cdot b}_{\ell m}$ modes of the spin-squared map are not generally zero when $B^a_{\ell m}=B^b_{\ell m}=0$.  Now we calculate explicitly the various skew-spectra
\begin{equation}\label{eq:C_lofEaEbc}
\begin{split}
    &C_{\ell}^{E^a E^{b\cdot c }} = -\frac{1}{2}(-1)^{s_b+s_c}\sum_{\ell_1\ell_2} \biggl[\mathcal{B}^{E^a E^{b\cdot c}}_{\ell\ell_1 \ell_2 }+(-1)^{\ell_1+\ell_2+\ell}\left(\mathcal{B}^{E^a E^{b\cdot c}}_{\ell\ell_1 \ell_2 }\right)^*
  \biggl]\sqrt{\frac{(2\ell_1+1)(2\ell_2+1)}{4\pi(2\ell+1)}}
  \begin{pmatrix}
      \ell_1 & \ell_2 & \ell\\
        -s_b & -s_c & (s_b+s_c)  
  \end{pmatrix},
\end{split}
\end{equation}
\begin{equation}\label{eq:C_lofEaBbc}
\begin{split}
    &C_{\ell}^{E^a B^{b\cdot c }} = -\frac{1}{2i}(-1)^{s_b+s_c}\sum_{\ell_1\ell_2} \biggl[\mathcal{B}^{E^a B^{b\cdot c}}_{\ell\ell_1 \ell_2 }-(-1)^{\ell_1+\ell_2+\ell}\left(\mathcal{B}^{E^a B^{b\cdot c}}_{\ell \ell_1 \ell_2}\right)^*
  \biggl]\sqrt{\frac{(2\ell_1+1)(2\ell_2+1)}{4\pi(2\ell+1)}}\begin{pmatrix}
      \ell_1 & \ell_2 & \ell\\
        -s_b & -s_c & (s_b+s_c)  
  \end{pmatrix},
\end{split}
\end{equation}
\begin{equation}\label{eq:C_lofBaEbc}
\begin{split}
    &C_{\ell}^{B^a E^{b\cdot c }} = -\frac{1}{2}(-1)^{s_b+s_c}\sum_{\ell_1\ell_2} \biggl[\mathcal{B}^{B^a E^{b\cdot c}}_{\ell\ell_1 \ell_2 }+(-1)^{\ell_1+\ell_2+\ell}\left(\mathcal{B}^{B^a E^{b\cdot c}}_{\ell\ell_1 \ell_2 }\right)^*
  \biggl]\sqrt{\frac{(2\ell_1+1)(2\ell_2+1)}{4\pi(2\ell+1)}}\begin{pmatrix}
      \ell_1 & \ell_2 & \ell\\
        -s_b & -s_c & (s_b+s_c)  
  \end{pmatrix},
\end{split}
\end{equation}
\begin{equation}\label{eq:C_lofBaBbc}
\begin{split}
    &C_{\ell}^{B^a B^{b\cdot c }} = -\frac{1}{2i}(-1)^{s_b+s_c}\sum_{\ell_1\ell_2} \biggl[\mathcal{B}^{B^a B^{b\cdot c}}_{\ell\ell_1 \ell_2 }-(-1)^{\ell_1+\ell_2+\ell}\left(\mathcal{B}^{B^a B^{b\cdot c}}_{\ell\ell_1 \ell_2 }\right)^*
  \biggl]\sqrt{\frac{(2\ell_1+1)(2\ell_2+1)}{4\pi(2\ell+1)}}\begin{pmatrix}
      \ell_1 & \ell_2 & \ell\\
        -s_b & -s_c & (s_b+s_c)  
  \end{pmatrix},
\end{split}
\end{equation}
where we have defined
\begin{equation}\label{eq:bispectrum_linear_comb}
    \mathcal{B}^{E^a E^{b\cdot c}}_{\ell\ell_1 \ell_2 } \equiv \mathcal{B}^{E^a B^{b\cdot c}}_{\ell\ell_1 \ell_2 } \equiv \left(\mathcal{B}^{E^a E^b E^c}_{\ell \ell_1\ell_2}-\mathcal{B}^{E^a B^b B^c}_{\ell \ell_1\ell_2}\right)+i\left(\mathcal{B}^{E^a E^b B^c}_{\ell \ell_1\ell_2}+\mathcal{B}^{E^a B^b E^c}_{\ell \ell_1\ell_2}\right),
\end{equation}
\begin{equation}\label{eq:bispectrum_linear_comb2}
    \mathcal{B}^{B^a E^{b\cdot c}}_{\ell_1 \ell_2 \ell} \equiv \mathcal{B}^{B^a B^{b\cdot c}}_{\ell_1 \ell_2 \ell} \equiv  \left(\mathcal{B}^{B^a E^b E^c}_{\ell \ell_1\ell_2}-\mathcal{B}^{B^a B^b B^c}_{\ell \ell_1\ell_2}\right)+i\left(\mathcal{B}^{B^a E^b B^c}_{\ell \ell_1\ell_2}+\mathcal{B}^{B^a B^b E^c}_{\ell \ell_1\ell_2}\right).
\end{equation}
\end{widetext}
We have used the definition of the isotropic angular bispectra Eq.~(\ref{eq:full_bispectrum}), the orthogonality of the Wigner-$3j$ Eq.~(\ref{eq:orthogonal}), and finally the definition of the angular skew-spectra Eq.~(\ref{eq:cross_power}). We note that we must impose a scale cut by taking the domain of the sum to be $0\leq\ell_1^{\rm{min}}\leq \ell_1 \leq \ell_1^{\rm{max}}$, $0\leq\ell_2^{\rm{min}}\leq \ell_2 \leq \ell_2^{\rm{max}}$, to avoid summing over (arbitrarily) small scales which are not accurately modeled. This is done in practice by filtering the fields present in the composed quantity in harmonic space before forming the composed quantity to remove these scales in the summation. See Ref.~\cite{Harscouet2025Fast} for a detailed description of this process and the practical impacts on it of a complex survey geometry. In principle we could consider various filters built to specifically select certain bispectrum configurations. We do not explore in detail this possibility, but note that it could be of interest for future work. 

One interesting feature of these statistics is the way the angular bispectra are combined to form the angular power spectra. We highlight that there is a complex sum of types of bispectra which enter into the power spectra. We note that the $C_{\ell}^{E^aE^{b\cdot c}}$ statistics have contributions from even parity bispectra only (see Eq.~(\ref{eq:bispectrum_parity}) for the definition of parity even and odd bispectra), since for $\ell_1+\ell_2+\ell =\rm{even}$, only the real part of Eq.~(\ref{eq:bispectrum_linear_comb}) contributes (which has bispectra with an even number of $B$ modes), and for $\ell_1+\ell_2+\ell =\rm{odd}$, only the imaginary part of Eq.~(\ref{eq:bispectrum_linear_comb}) contributes (which consists of bispectra with an odd number of $B$ modes). The opposite is true of $C_{\ell}^{E^aB^{b\cdot c}}$, which only has contributions from odd parity bispectra. Similarly, $C_{\ell}^{B^aE^{b\cdot c}}$  and $C_{\ell}^{B^aB^{b\cdot c}}$ receive contributions from the odd parity and even parity bispectra respectively. These results can be seen quickly by noting that we have constructed power spectra which retain the parity structure 
\begin{equation}
    \mathbb{P} \left(C_{\ell}^{F^a G^{b\cdot c }}\right) = P_{FG}C_{\ell}^{F^a G^{b\cdot c }},
\end{equation}
where $P_{EE}=P_{BB}=1$ and $P_{EB}=P_{BE}=-1$ (compare to Eq.~(\ref{eq:power_parity})).

It is tempting to think that there might be a general way to linearly combine the four statistics for three different fields to extract the contribution from one type of bispectrum for general filters. We see that in general (i.e. for three different fields and arbitrary filters) Eqs. (\ref{eq:C_lofEaEbc}), (\ref{eq:C_lofEaBbc}), (\ref{eq:C_lofBaEbc}) and (\ref{eq:C_lofBaBbc}) alone are not able to be linearly combined to isolate the individual contributions. This is because each type of bispectrum appears in a sum with one other type. We could consider finding all the power spectra of the form $F^a\times G^{b^*\cdot c}$ (i.e. $EE$, $EB$, $BE$ and $BB$), which will yield new statistics with new linear combinations of the bispectra, but the Wigner-$3j$ involved in the summation being different ($s_b\rightarrow -s_b$) means that it is not possible in general to isolate the individual bispectrum types via linear combinations\footnote{One might think to try finding all the power spectra of $F^{(a^*)}\times G^{b\cdot c}$ for example, but as pointed out in Sec.~\ref{sec:general_formalism}, this is proportional to the previous statistics (this makes sense since the information contained in $a$ should be the same as $a^*$).}. We describe another approach to deal with this issue in Appendix~\ref{sec:appendix_eb_squared}.

We can also consider statistics where two of the three fields are the same. For $F^a \times G^{a\cdot b}$ (or $F^a \times G^{a^*\cdot b}$) we find that there are $4$ independent equations with $6$ types of bispectra for a given $\ell$ space configuration (even or odd). We have more freedom than previously, meaning we can form statistics with new linear combinations of bispectra. For example, we can construct statistics dependent on linear combinations such as $\mathcal{B}^{E^a E^a E^b}_{\ell \ell_1\ell_2}+\mathcal{B}^{B^a B^a E^b}_{\ell \ell_1\ell_2}$ and $\mathcal{B}^{B^a B^a B^b}_{\ell \ell_1\ell_2}+\mathcal{B}^{E^a E^a B^b}_{\ell \ell_1\ell_2}$. For $F^a\times G^{b^2}$ we are able to construct two statistics which depend only on one type of bispectrum, $\mathcal{B}^{E^aE^bB^b}_{\ell \ell_1\ell_2}$ and $\mathcal{B}^{B^aE^bB^b}_{\ell \ell_1\ell_2}$ respectively. We also note that the $F^a\times G^{b^*\cdot b}$ statistics have the special property that the imaginary parts of Eqs.~(\ref{eq:bispectrum_linear_comb}) and (\ref{eq:bispectrum_linear_comb2}) are zero. For the statistics constructed from the same three fields of the form $F^a\times G^{a^2}$, it is possible to construct statistics where only one type of bispectrum contributes. We see for example that in Eq.~(\ref{eq:bispectrum_linear_comb}) the imaginary part reduces to $\mathcal{B}^{E^a E^a B^a}_{\ell \ell_1\ell_2}+\mathcal{B}^{E^a B^a E^a}_{\ell \ell_1\ell_2}=2\mathcal{B}^{E^a E^a B^a}_{\ell \ell_1\ell_2}$. We are able to do a Gaussian elimination to fully identify each of the contributions. We note that this is not achievable for $F^a\times G^{a^* \cdot a}$. We emphasize that specific filters which select only one bispectrum configuration will circumvent the problems discussed here, but will no longer include a sum over bispectra. \\

\section{Application to weak lensing}\label{sec:weak_lensing_spin_fields}
\subsection{Theory}
In this section we apply the formalism developed in Sec~\ref{sec:spin-squared_maps} to weak lensing, i.e. the distortion of light by the large scale structure of the universe, and its cross-correlation with projected galaxy clustering. Weak lensing is quantified by the linear mapping of the image plane to the celestial sphere described by the $2\times2$ distortion matrix, $\boldsymbol{\mathcal{H}}(\nv)$. We can separate out the physical effects contributing to the distortion of the image by decomposing this tensor under $U(1)$. We define the basis connecting the $2D$ real quantity (tensor field) to the $1D$ complex quantity (spin-$s$ field) as follows
\begin{equation}
    \boldsymbol{e}_{\pm}\equiv \boldsymbol{e}_{\hat{\theta}}\pm i \boldsymbol{e}_{\hat{\varphi}},
\end{equation}
where the unit vectors 
\begin{equation}
\begin{split}
    \boldsymbol{e}_{\hat{\theta}}&\equiv (\cos\theta \cos\phi, \cos\theta\sin\phi,-\sin\theta),\\
    \boldsymbol{e}_{\hat{\varphi}} &\equiv (-\sin\phi,\cos\phi,0),
\end{split}
\end{equation}
are spherical polar coordinates. The irreducible parts of the distortion tensor can then be found to be 
\begin{equation}
    \gamma(\nv) \equiv \mathcal{H}^{ij} \boldsymbol{e}_{+,i}\boldsymbol{e}_{+,j} = \mathcal{H}^{11} - \mathcal{H}^{22} + i\left(\mathcal{H}^{12}+\mathcal{H}^{21}\right), 
\end{equation}
and
\begin{equation}
    \zeta(\nv)  \equiv \mathcal{H}^{ij} \boldsymbol{e}_{+,i}\boldsymbol{e}_{-,j} = \mathcal{H}^{11} + \mathcal{H}^{22} + i\left(\mathcal{H}^{12}-\mathcal{H}^{21}\right),
\end{equation}
where $\gamma\equiv \gamma_1+i\gamma_2$ is the spin-$2$ object describing the shear of the image, and $\zeta \equiv \kappa + i \omega$ is the spin-$0$ object formed by $\kappa$ and $\omega$, which describe, respectively, the convergence and rotation of the image. Consider forming the tensor product of the distortion tensor with itself to construct the four-indexed object
\begin{equation}
    \boldsymbol{\mathcal{E}}(\nv) = \boldsymbol{\mathcal{H}}(\nv) \otimes \boldsymbol{\mathcal{H}}(\nv),
\end{equation}
where the components are 
\begin{equation}
    \mathcal{E}^{ijkl}(\nv) \equiv \mathcal{H}^{ij}(\nv)\  \mathcal{H}^{kl}(\nv).
\end{equation}
We decompose this rank-$4$ tensor into its irreducible parts under $U(1)$ to find the physical quantities of interest. We have 1 spin-$4$ field (which comes with an associated spin-$(-4)$ field), with two real degrees of freedom:
\begin{equation}
    \gamma^2 \equiv \mathcal{E}^{ijkl}\boldsymbol{e}_{+,i}\boldsymbol{e}_{+,j}\boldsymbol{e}_{+,k}\boldsymbol{e}_{+,l} =  \gamma_1^{2}-\gamma_2^2 + 2i\gamma_1\gamma_2.
\end{equation}
We have two spin-$2$ fields (each with associated spin-$(-2)$ fields) which are
\begin{equation}
    \gamma \zeta = \kappa \gamma_1 -\omega \gamma_2 + i(\omega\gamma_1+\gamma_2\kappa),
\end{equation}
and
\begin{equation}
    \gamma \zeta^* = \kappa\gamma_1 + \omega\gamma_2 + i (\gamma_2 \kappa - \gamma_1\omega).
\end{equation}
Finally, we find that we have 3 spin-$0$ terms
\begin{equation}
    \zeta^2 = \kappa^2 -\omega^2 +2i\kappa\omega,
\end{equation}
\begin{equation}
    \zeta\zeta^* = \kappa^2 +\omega^2,
\end{equation}
\begin{equation}
    \gamma\gamma^* = \gamma_1^2+\gamma_2^2.
\end{equation}
In effect, we have constructed all the possible independent quadratic combinations of the irreducible components of the distortion tensor, which is exactly equivalent to considering all the independent products of the spin fields, as described in Sec.~\ref{sec:spin-squared_maps}. We have omitted the redshift bin indices for simplicity.

We can also consider forming cross-correlations with galaxy clustering in $2D$, which is related to the $3D$ quantity as
\begin{equation}\label{eq:proj_density}
\delta_g^{i} (\nv) = \int\mathrm{d}z\ p_{i}(z) \delta_g (\chi(z) \nv, z) 
\end{equation}
where $p_i(z)$ is the redshift distribution of galaxies in a given redshift bin $i$. We then form 
\begin{equation}
    \boldsymbol{\mathcal{H}}'(\nv) = \delta_g \boldsymbol{\mathcal{H}}(\nv),
\end{equation}
which decomposes into the spin-$2$ object
\begin{equation}
    \delta_g\gamma = \delta_g\gamma_1 + i\delta_g\gamma_2,
\end{equation}
and the spin-$0$ quantity
\begin{equation}
    \delta_g\zeta = \delta_g\kappa +i\delta_g\omega.
\end{equation}

We have kept the formalism relatively general up to now. In practice we write down expressions for the weak lensing quantities at linear order (see Refs.\cite{Bernardeau:2009bm,Bernardeau_2012} for derivations) using the scalar perturbed Friedmann-Lema\^itre-Robertson-Walker (FLRW) metric and the Born approximation to find
\begin{equation}\label{eq:shear_field}
    \gamma^i = \int \mathrm{d}z \ p_i(z)\frac{1}{2}\eth^2 \Phi_L,
\end{equation}
\begin{equation}
    \zeta^i = \int \mathrm{d}z \ p_{i}(z)\bar{\eth}\eth \Phi_L,
\end{equation}
where
\begin{equation}
\begin{split}
    &\Phi_L(\hat{n}_0,\chi_s)  \equiv 2\int_{0}^{\chi_s}\mathrm{d}\chi \frac{\chi_s - \chi}{\chi_s\chi} \Psi\\
    &= 2\int_{0}^{\chi_s}\mathrm{d}\chi \frac{\chi_s - \chi}{\chi_s\chi} \left(\int \frac{\mathrm{d}^3 k}{(2\pi)^3}\frac{4\pi G a^2 \bar{\rho}}{k^2}\delta(\boldsymbol{k},z(\chi)) e^{i\chi \bf k \cdot \nv }\right),
\end{split}
\end{equation}
is the lensing potential, constructed from the Weyl potential $\Psi$, which we relate to the matter overdensity, $\delta=\bar{\rho}/\rho-1$, using Poisson's equation in the second line, where $a$ is the scale factor. We have assumed photon propagation in a flat space-time described by General Relativity. We note that at linear order we do not have a rotation component coming from $\omega$, and the $B$-modes of $\gamma$ are zero. We will assume this from now on.

In the context of studying the large scale structure, we are currently only able to measure the so-called reduced shear 
\begin{equation}
     \frac{\gamma}{1-\kappa} \approx \gamma + \mathcal{O}(\kappa\gamma).
\end{equation}
We consider weak lensing, where $\kappa\ll1$, and focus on $\gamma$. We highlight that this does not involve any mass-mapping process such as the Kaiser-Squires algorithm. We do not take into account corrections to the reduced shear, intrinsic alignments, or magnification bias. We shall concentrate on making theoretical predictions for the statistics of the shear and the correlations with $2D$ galaxy clustering in this simplified scenario. In particular, we focus on the quadratic quantities: $\gamma^2$, $\gamma^*\gamma$, $\delta_g\gamma$ and $\delta_g^2$. 

The angular statistics can be shown to be related to their $3$D counterparts. We define the $3$D power spectrum as
\begin{equation}
    \langle r(\boldsymbol{k}_1)s(\boldsymbol{k}_2)\rangle = (2\pi)^3\delta^{(3)}(\boldsymbol{k}_1-\boldsymbol{k}_2)P^{rs}(k_1),
\end{equation}
and the bispectrum 
\begin{align}\nonumber
    &\langle r(\boldsymbol{k}_1)s(\boldsymbol{k}_2)t(\boldsymbol{k}_3)\rangle = \\
    &\hspace{30pt}(2\pi)^3\delta^{(3)}(\boldsymbol{k}_1+\boldsymbol{k}_2+\boldsymbol{k}_3)B^{rst}(k_1, k_2,k_3),
\end{align}
where $r,s,t$ represent arbitrary $3D$ homogeneous cosmological fields. We further define the Fourier transform as 
\begin{equation}
    r(\boldsymbol{k}) = \int\mathrm{d}^3x\ r(\boldsymbol{x})e^{i\boldsymbol{k}\cdot \boldsymbol{x}}.
\end{equation}
Using these definitions, we find in the Limber approximation (valid for large $\ell$)\footnote{The Limber approximation \cite{LoVerde_2008, Kilbinger_2017,Lemos_2017} makes use of the fact that the projection kernel is smooth and wide on small angular scales compared to the $3D$ quantity that is being projected onto the celestial sphere. This amounts to the replacement of $j_\ell(k\chi)\rightarrow \sqrt{\pi/{(2\ell+1)}}\delta^{\cal D}\left (k\chi - \ell-1/2  \right)$ inside the integrals, where $j_\ell$ is a Bessel function. }, 
\begin{equation}
    C_{\ell}^{ab,ij} = \int \frac{\mathrm{d}\chi}{\chi^2} q^a_i(\chi)q^b_{j}(\chi)P^{\delta\delta}\left(k=\frac{\ell+1/2}{\chi},z(\chi)\right),
\end{equation}
and
\begin{equation}\label{eq:projected_bispectra}
\begin{split}
    b_{\ell_1\ell_2\ell_3}^{abc,ijk} &= \int \frac{\mathrm{d}\chi}{\chi^4} q^a_{i}(\chi)q^b_{j}(\chi)q^c_{k}(\chi) \times \\
    &\times B^{\delta\delta\delta}\left(\frac{\ell_1+1/2}{\chi},\frac{\ell_2+1/2}{\chi},\frac{\ell_3+1/2}{\chi},z(\chi)\right),
\end{split}
\end{equation}
where the cosmological fields $a,b,c$ are associated with multipoles $\ell_1$, $\ell_2$, $\ell_3$ and redshift bins $i,j,k$ respectively. We consider here two types of fields: galaxy clustering $\delta_g(\nv)$ and the $E$-mode of the shear, $E^\gamma(\nv)$. We can show using Eqs.~\ref{eq:proj_density} and  \ref{eq:shear_field} that these are associated with the kernels
\begin{equation}\label{eq:galaxy_kernel}
    q^g_i(\chi)  = H(\chi)p_{i}(\chi)
\end{equation}
and 
\begin{equation}\label{eq:lensing_kernel}
    q^\gamma_i(\chi)  =G_\ell \frac{3H_0^2\Omega_M}{2a(\chi)c} \chi \int_{z(\chi)}^{\infty}\mathrm{d}z'p_{i}(z')\left(1-\frac{\chi}{\chi(z')}\right),
\end{equation}
respectively where
\begin{equation}
    G_\ell =    \sqrt{\frac{(\ell+2)!}{(\ell-2)!}}\frac{1}{(\ell+1/2)^2}.
\end{equation}
Note that $G_\ell$ is associated with  the multipole, $\ell$, of the $\gamma$ field and is negligibly close to $1$ for $\ell \geq 10$.  We often denote the $\delta_g(\nv)$ field as $g$ for ease of notation. We have used the definition of the reduced bispectrum of Eq.~(\ref{eq:reduced_bispectrum}) in Eq.~(\ref{eq:projected_bispectra}) since only the parity-even angular bispectra are non-zero in this approximation. Using the results from Sec.~\ref{sec:spin-squared_maps}, the quantities we consider in this paper, which could be of interest for future analyses, are the following:
\begin{equation}\label{eq:ClEEgammasquared_lensing}
\begin{split}
   C^{E^{\gamma} \times E^{\gamma^2}}_{\ell} &=-\sum_{\ell_1\ell_2} \frac{(2\ell_1+1)(2\ell_2+1)}{4\pi}
  \begin{pmatrix}
      \ell_1 & \ell_2 & \ell\\
        0 & 0 & 0 
  \end{pmatrix}\times\\
  &\times\begin{pmatrix}
      \ell_1 & \ell_2 & \ell\\
        -2 & -2 & 4  
  \end{pmatrix}b^{E^{\gamma} E^{\gamma}E^{\gamma}}_{\ell \ell_1 \ell_2 },
\end{split}
\end{equation}
\begin{equation}\label{eq:ClEEgammastar_lensing}
\begin{split}
   C^{E^{\gamma} \times E^{\gamma^*\gamma}}_{\ell} &=-\sum_{\ell_1\ell_2} \frac{(2\ell_1+1)(2\ell_2+1)}{4\pi}
  \begin{pmatrix}
      \ell_1 & \ell_2 & \ell\\
        0 & 0 & 0 
  \end{pmatrix} \times \\
  &\times 
  \begin{pmatrix}
      \ell_1 & \ell_2 & \ell\\
        -2 & 2 & 0  
  \end{pmatrix}b^{E^{\gamma} E^{\gamma}E^{\gamma}}_{\ell \ell_1 \ell_2 },
\end{split}
\end{equation}
\begin{equation}\label{eq:galaxy_gamma_squared}
    \begin{split}
   C^{g \times E^{\gamma^2}}_{\ell} &=-\sum_{\ell_1\ell_2} \frac{(2\ell_1+1)(2\ell_2+1)}{4\pi}
  \begin{pmatrix}
      \ell_1 & \ell_2 & \ell\\
        0 & 0 & 0 
  \end{pmatrix}\times\\
  &\times\begin{pmatrix}
      \ell_1 & \ell_2 & \ell\\
        -2 & -2 & 4  
  \end{pmatrix}b^{g E^{\gamma}E^{\gamma}}_{\ell \ell_1 \ell_2 },
\end{split}
\end{equation}
\begin{equation}\label{eq:galaxy_gamma_gammastar}
\begin{split}
   C^{g \times E^{\gamma^*\gamma}}_{\ell} &=-\sum_{\ell_1\ell_2} \frac{(2\ell_1+1)(2\ell_2+1)}{4\pi}
  \begin{pmatrix}
      \ell_1 & \ell_2 & \ell\\
        0 & 0 & 0 
  \end{pmatrix} \times \\
  &\times 
  \begin{pmatrix}
      \ell_1 & \ell_2 & \ell\\
        -2 & 2 & 0  
  \end{pmatrix}b^{g E^{\gamma}E^{\gamma}}_{\ell \ell_1 \ell_2 },
\end{split}
\end{equation}
\begin{equation}\label{eq:gamma_galaxy_galaxy}
\begin{split}
   C^{E^{\gamma} \times g^2}_{\ell} &=\sum_{\ell_1\ell_2} \frac{(2\ell_1+1)(2\ell_2+1)}{4\pi}
  \begin{pmatrix}
      \ell_1 & \ell_2 & \ell\\
        0 & 0 & 0 
  \end{pmatrix}^2 b^{E^{\gamma} gg}_{\ell \ell_1 \ell_2 },
\end{split}
\end{equation}
\begin{equation}\label{eq:gamma_gamma_galaxy}
\begin{split}
   C^{E^{\gamma} \times E^{\gamma g}}_{\ell} &=\sum_{\ell_1\ell_2} \frac{(2\ell_1+1)(2\ell_2+1)}{4\pi}
  \begin{pmatrix}
      \ell_1 & \ell_2 & \ell\\
        0 & 0 & 0 
  \end{pmatrix}\times \\
  &\times \begin{pmatrix}
      \ell_1 & \ell_2 & \ell\\
        -2 & 0 & 2 
  \end{pmatrix}b^{E^{\gamma} E^{\gamma}g}_{\ell \ell_1 \ell_2 },
\end{split}
\end{equation}
\begin{equation}\label{eq:galaxy_gamma_galaxy}
\begin{split}
   C^{g \times E^{\gamma g}}_{\ell} &=\sum_{\ell_1\ell_2} \frac{(2\ell_1+1)(2\ell_2+1)}{4\pi}
  \begin{pmatrix}
      \ell_1 & \ell_2 & \ell\\
        0 & 0 & 0 
  \end{pmatrix}\times \\
  &\times \begin{pmatrix}
      \ell_1 & \ell_2 & \ell\\
        -2 & 0 & 2 
  \end{pmatrix}b^{gE^{\gamma}g}_{\ell \ell_1 \ell_2 },
\end{split}
\end{equation}
\begin{equation}\label{eq:ggg}
\begin{split}
   C^{g \times g^2}_{\ell} &=\sum_{\ell_1\ell_2} \frac{(2\ell_1+1)(2\ell_2+1)}{4\pi}
  \begin{pmatrix}
      \ell_1 & \ell_2 & \ell\\
        0 & 0 & 0 
  \end{pmatrix}^2b^{ggg}_{\ell \ell_1 \ell_2 },
\end{split}
\end{equation}
where we have suppressed the redshift bins for simplicity. Note that the parity-even condition, in addition to approximating the $B$-modes to be zero (in practice this might produce a small correction to the signal), means that the $C_{\ell}^{E^aB^{b\cdot c}}$ power spectra in Eq.~(\ref{eq:C_lofEaBbc}) vanish. As expected, we see that the quantities differ by a Wigner-$3j$ symbol and sometimes a minus sign (coming from considering $g$ instead of $E^g \equiv -g$). We emphasize again that in practice the fields must be filtered before forming quadratic quantities to avoid contributions from arbitrarily small scales.
\begin{figure*}
    \centering 
    \includegraphics[width=1\linewidth]{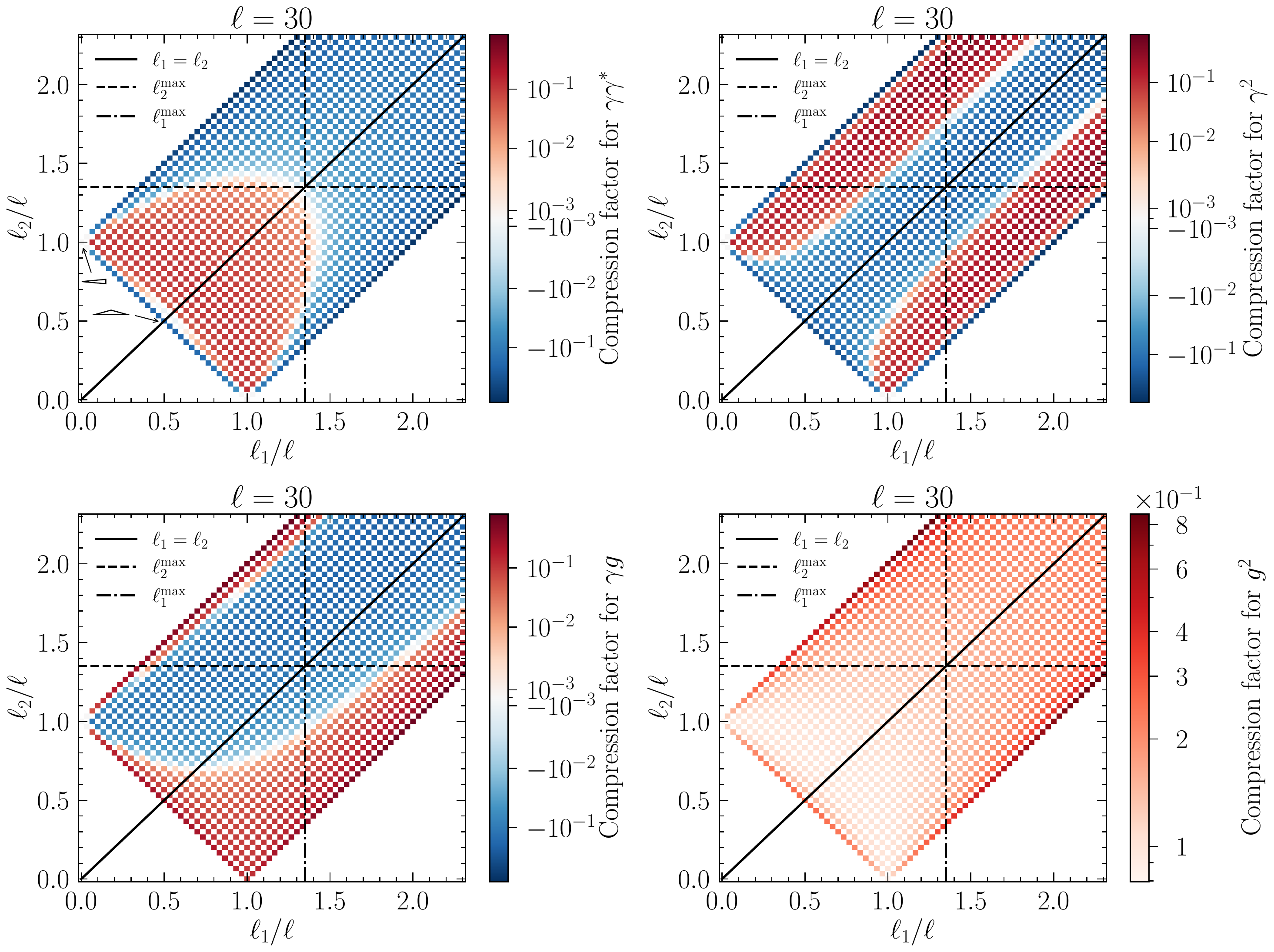}
    \caption{The compression factors present in Eqs.~(\ref{eq:ClEEgammasquared_lensing})-(\ref{eq:ggg}) (these are the factors multiplied by the angular bispectrum at a given $(\ell,\ell_1,\ell_2)$ configuration before summing over $\ell_1$ and $\ell_2$, where we only consider $\ell_1+\ell_2+\ell=\text{even}$ contributions) plotted for $\ell_1/\ell$ against $\ell_2/\ell$, all shown at a slice of $\ell=30$. The top left (right) plot corresponds to the compression factors related to $\gamma^*\gamma$ ($\gamma^2$) shown in Eqs.~(\ref{eq:ClEEgammastar_lensing}) and (\ref{eq:galaxy_gamma_gammastar}) (Eqs.~(\ref{eq:ClEEgammasquared_lensing}) and (\ref{eq:galaxy_gamma_squared})). The bottom left (right) plot shows the compression factors related to $\gamma g$ ($g^2$) shown in Eqs.~(\ref{eq:gamma_gamma_galaxy}) and (\ref{eq:galaxy_gamma_galaxy})   (Eqs.~(\ref{eq:gamma_galaxy_galaxy}) and (\ref{eq:ggg})). The squeezed ($\ell_1/\ell \ll \ell_2/\ell$) and folded ($\ell_1/\ell=\ell_2/\ell=1/2$) configurations are illustrated in the top left plot at $(0,1)$ and $(0.5,0.5)$ respectively. We show on the plots a choice of $\ell_1^{\rm{max}}=40$ and $\ell_2^{\rm{max}}=40$ to demonstrate the effect of filtering the domain of $\ell_1,\ell_2$, as the regions below and to the left of these lines contribute to the angular power spectrum at a given $\ell$.}
    \label{fig:wigner_coefficient}
\end{figure*}

\subsection{Predictions for $\Lambda$CDM}
We will now explore the form of these various statistics in $\Lambda$CDM. We choose the fiducial cosmology to match the Planck 2018 results \cite{planckresults}, i.e. $\{\sum m_\nu ,\Omega_b,\Omega_c,h,n_s,\sigma_8\}= \{0.06,0.0493,0.264,0.6766,0.9665,0.8102 \}$, where $\sum m_\nu$ is the sum of the neutrino masses, $\Omega_b$ and $\Omega_c$ are the baryon and cold dark matter energy density fractions respectively, $h$ is the Hubble constant in $100 \ \rm{km} \ \rm{s}^{-1} \rm{Mpc}^{-1}$, $n_s$ is the tilt of the primordial power spectrum and $\sigma_8$ describes the root mean square of the amplitude of matter perturbations on scales of $8h^{-1}\rm{Mpc}$. First consider the angular bispectrum, from which the statistics are constructed. We make theoretical predictions for the $3D$ matter bispectrum using the fitting function BiHalofit \cite{Takahashi_2020}, able to describe small, non-linear scales. The model recovers the $3D$ bispectrum from $N$-body simulations to an accuracy of $20 \%$ for $w\text{CDM}$ models, where $w$ refers to a constant dark energy equation of state, on scales of $k < 3\ h \ \rm{Mpc}^{-1}$ in the redshift range of $z=0$ to $z=1.5$. The authors find that the convergence bispectra were similarly accurate at $\ell \lesssim 4000$ for a source redshift of $z=1$.

We compute the kernels of  Eqs.~(\ref{eq:galaxy_kernel}) and (\ref{eq:lensing_kernel}) using the Core Cosmology Library (CCL) \cite{Chisari_2019}. We consider two Gaussian distributed redshift bins for simplicity. One for the shear lensing of $z_\gamma \sim \mathcal{N}(\bar{z}=0.5,\sigma_z=0.05)$ and a nearer redshift bin for the galaxy kernel $z_g \sim \mathcal{N}(\bar{z}=0.25,\sigma_z=0.05)$. We note that it is possible to form statistics with up to three different redshift bins. We also choose a linear galaxy bias of $1$, i.e. $\delta_g\sim b \delta = \delta$ \cite{Desjacques_2018}. 

The angular bispectra we consider here exhibit a strong dependence on triangle configuration \cite{Munshi_2020}. In particular, the squeezed shapes, with $\ell_1\ll \ell_2= \ell$, generally produce the largest amplitudes across most multipoles. The folded configurations, $\ell_1=\ell_2=\ell/2$, typically yield the next-largest signals, while the equilateral shapes, $\ell_1=\ell_2=\ell$, tend to have the smallest amplitudes. We note that the amplitude of the bispectrum for the folded configuration is larger than the squeezed configurations for low $\ell$ and that larger triangles in general have a lower bispectrum signal. \\
\begin{figure*}
    \centering
    \includegraphics[width=1\linewidth]{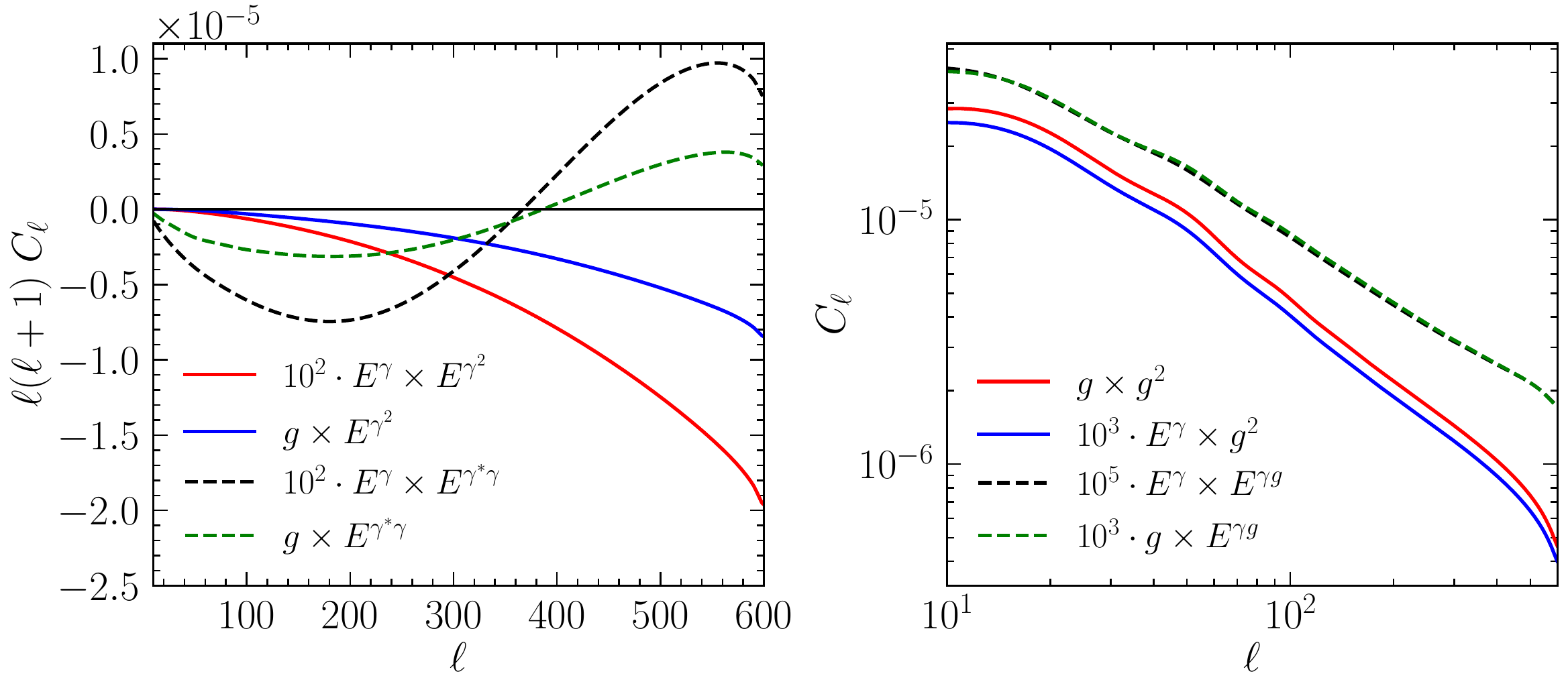}
    \caption{Theoretical predictions based on BiHalofit for the angular power spectra shown in Eqs.(\ref{eq:ClEEgammasquared_lensing})-(\ref{eq:ggg}) formed from the quadratic quantities: $\gamma^2$, $\gamma^*\gamma$, $\gamma g$ and $g^2$. The power spectra for Eqs.(\ref{eq:ClEEgammasquared_lensing})-(\ref{eq:galaxy_gamma_gammastar}), (${E^\gamma\times E^{\gamma^2}}$, ${E^\gamma\times E^{\gamma^*\gamma}}$, ${g\times E^{\gamma^2}}$, and ${g\times E^{\gamma^*\gamma}}$), are shown in the left-hand plot. The remaining statistics of Eqs.(\ref{eq:gamma_galaxy_galaxy})-(\ref{eq:ggg}) (${g\times g^2}$, ${E^\gamma\times g^2}$, ${\gamma \times E^{\gamma g}}$, and ${g\times E^{\gamma g}}$) are shown in the right-hand plot. All the statistics are shown for the fiducial Planck 2018 cosmology \cite{planckresults}, i.e. $\{\sum m_\nu ,\Omega_b,\Omega_c,h,n_s,\sigma_8\}= \{0.06,0.0493,0.264,0.6766,0.9665,0.8102 \}$, with redshift bins for the lensing sample of $z_\gamma \sim \mathcal{N}(\bar{z}=0.5,\sigma_z=0.05)$ and  for the galaxy sample of $z_g \sim \mathcal{N}(\bar{z}=0.25,\sigma_z=0.05)$. We neglect the contribution from $B$ modes and parity odd signals. We choose the domain of $\ell_1$ and $\ell_2$ to be and $[\ell_1^{\rm{min}}, \ell_1^{\rm{max}}]=[10,600]$ and $[\ell_2^{\rm{min}}, \ell_2^{\rm{max}}]=[10,600]$.  }
    \label{fig:angular_power}
\end{figure*}
\begin{figure*}
    \centering
    \includegraphics[width=1\linewidth]{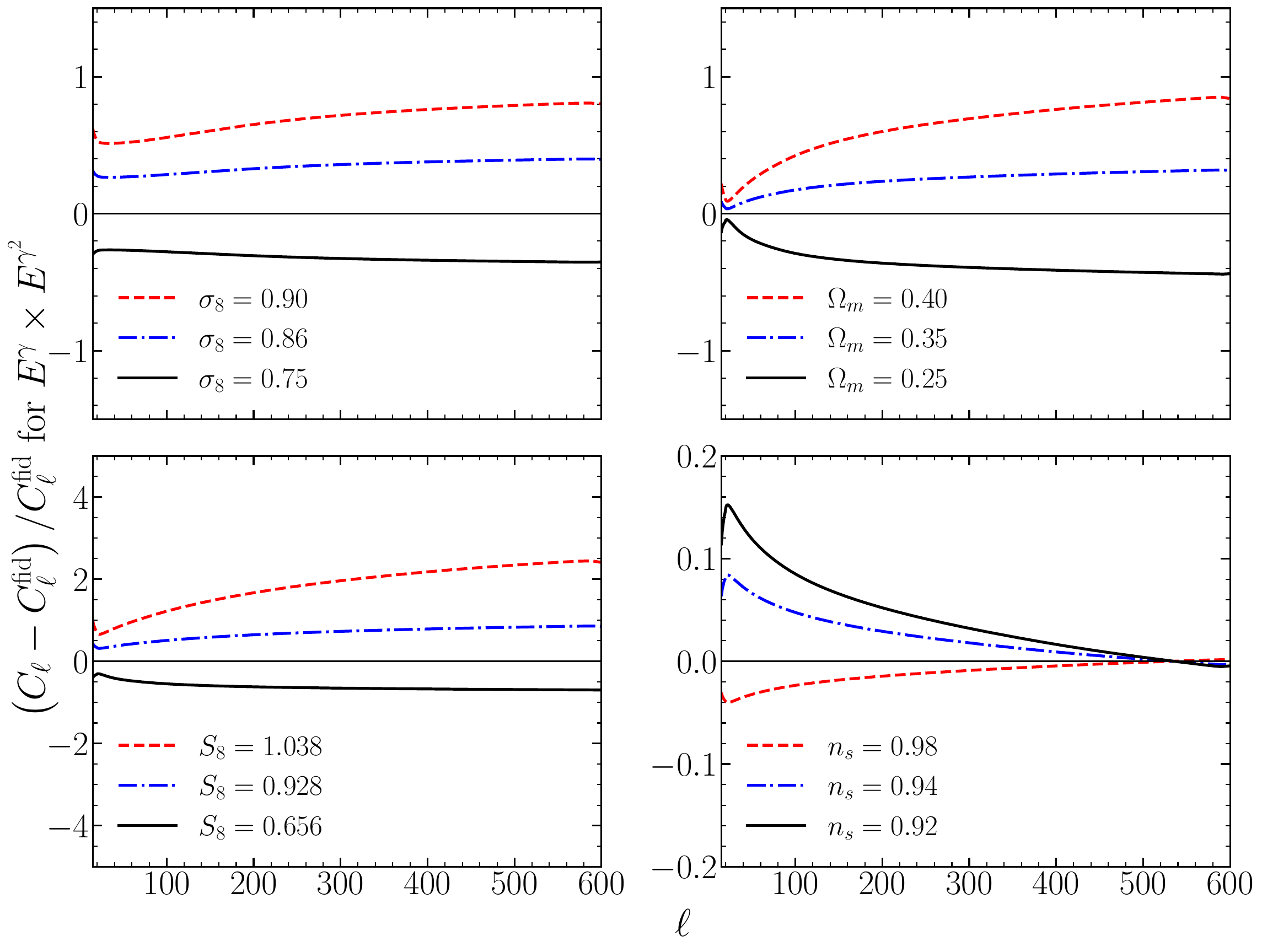}
    \caption{Ratio of power spectra for $E^\gamma\times E^{\gamma^2}$ of Eq.~(\ref{eq:ClEEgammasquared_lensing}) for different cosmologies relative to the fiducial Planck \cite{planckresults} results, taking $\{\sum m_\nu ,\Omega_b,\Omega_c,h,n_s,\sigma_8\}= \{0.06,0.0493,0.264,0.6766,0.9665,0.8102 \}$, with the same redshift bin of $z_\gamma \sim \mathcal{N}(\bar{z}=0.5,\sigma_z=0.05)$ for each of the shear quantities. We use the domain of the sum in the statistics of  $[\ell_1^{\rm{min}},\ell_1^{\rm{max}}]=[10,600]$ and $[\ell_2^{\rm{min}},\ell_2^{\rm{max}}]=[10,600]$.}
    \label{fig:gamma_squared_ratios_real_cosmologies}
\end{figure*}
\begin{figure*}
    \centering
    \includegraphics[width=1\linewidth]{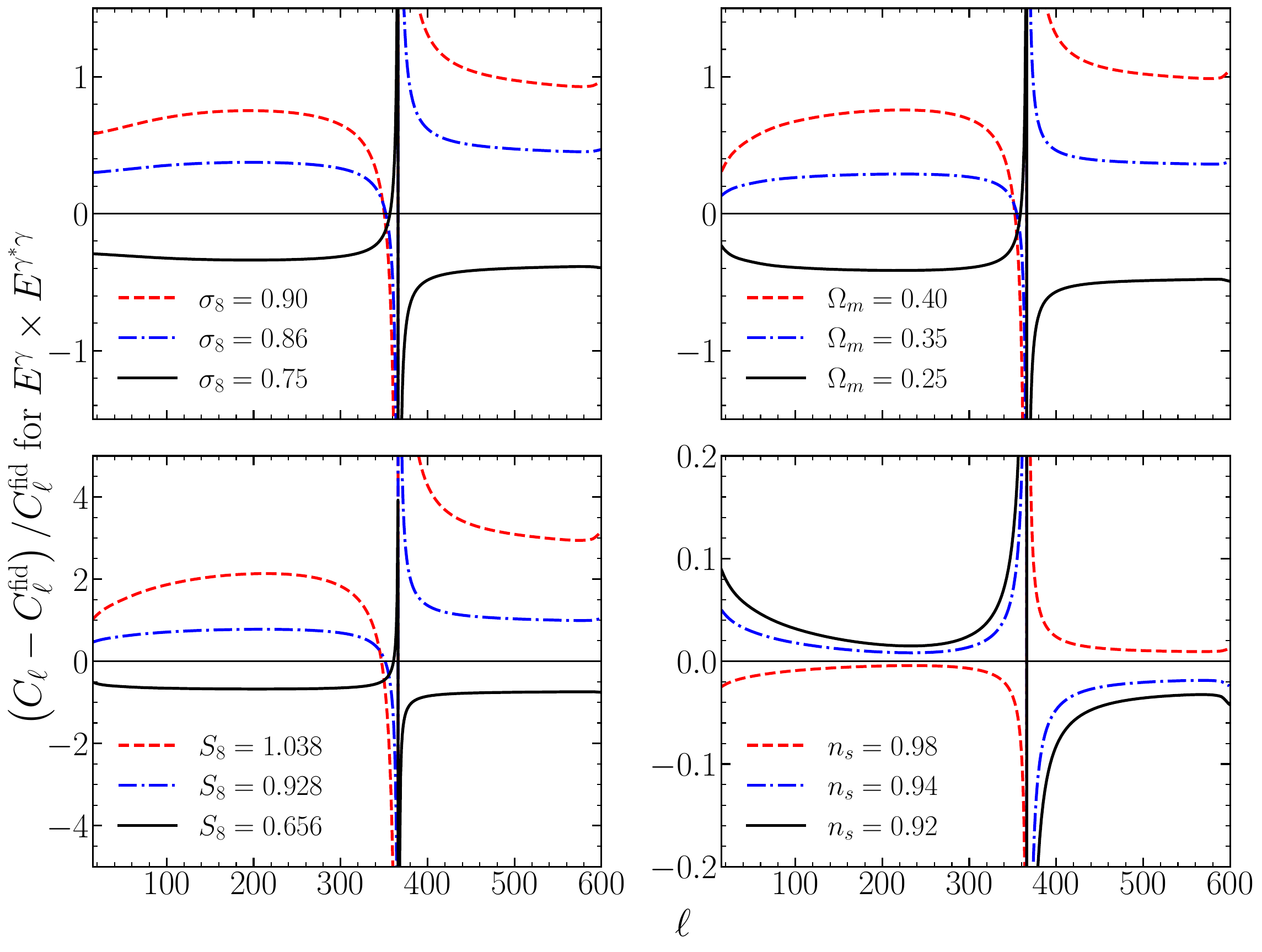}
    \caption{The same as Fig.~\ref{fig:gamma_squared_ratios_real_cosmologies} but for the ratio of power spectra for $E^\gamma\times E^{\gamma^*\gamma}$ of Eq.~(\ref{eq:ClEEgammastar_lensing}).}
    \label{fig:gamma_gammastar_ratios_real_cosmologies}
\end{figure*}
We turn now to the calculation of the angular power spectra, found by summing over the product of the angular bispectra with the Wigner-$3j$ coefficients in Eqs.~(\ref{eq:ClEEgammasquared_lensing})-(\ref{eq:ggg}). We show the shape-dependence of the Wigner-$3j$ factors in Fig.~\ref{fig:wigner_coefficient}, where the `compression factor' is the factor multiplied by the bispectrum before summing over $\ell_1$ and $\ell_2$. We see that the compression factors are non-zero only in certain regions of the plot. This reflects both the modeling (in working at linear order and therefore neglecting the contribution from $B$ modes) and the various conditions imposed on the bispectrum by isotropy and parity arguments. We only include $\ell_1+\ell_2+\ell=\text{even}$, meaning that the plane has a checkerboard-like pattern. The other regions are zero since the triangle conditions are not satisfied, $|\ell_1-\ell_2|\leq \ell \leq \ell_1+\ell_2$.  We find that the size of the compression factor for equilateral and squeezed configuration depends weakly on $\ell$, but depends strongly on $\ell$ for the isosceles and folded configurations, increasing in magnitude with increasing $\ell$. These observations are reflected in the general shape of the contour plots for other $\ell$ slices being found to be the same. The shape dependence of the angular power spectra preserves the same hierarchy seen in the angular bispectrum across most of the $\ell$ range: the squeezed configurations have the largest amplitudes, followed by the folded configurations, with the equilateral shapes remaining the smallest.

We can see that the choice of the domain of $\ell_1$ and $\ell_2$ to include in the summation will affect the size and shape of the angular power spectrum. We see that the sign of the region which is below the $\ell_1^{\rm{max}}$ and $\ell_2^{\rm{max}}$ lines will dictate the sign of the final angular power spectrum at that given $\ell$. For example, consider the $\gamma^*\gamma$ statistic. We see that for $\ell\ll
\ell_1^{\rm{max}},\ell_2^{\rm{max}}$, the summation will be over a largely negative region, which makes the angular power at that $\ell$ negative (despite the bispectrum amplitude varying over these scales). Making this range smaller typically  decreases the magnitude of the statistics.  We plot the angular power spectra of all the statistics in Fig.~\ref{fig:angular_power}, where we choose the domain of $\ell_1$ and $\ell_2$ to be $[\ell_1^{\rm{min}}, \ell_1^{\rm{max}}]=[10,600]$ and $[\ell_2^{\rm{min}}, \ell_2^{\rm{max}}]=[10,600]$ for simplicity and because we expect both the Limber approximation and BiHalofit to be accurate on these scales \cite{Takahashi_2020}. The statistics in the left hand plot are multiplied by $\ell(\ell+1)$ to more easily see the smaller scales. The plots show differences in the magnitude of the different statistics, where statistics including galaxy clustering are larger. We note that there are fictitious decreases and increases of the magnitude of the power on scales close to $\ell_1^{\rm{max}}$ and $\ell_2^{\rm{max}}$ (particularly visible in the left-hand plot of Fig.~\ref{fig:angular_power}). These can be easily interpreted by noting that we are excluding a region of the plots in Fig.~\ref{fig:wigner_coefficient} and the sign of the relevant region will dictate whether the power spectra are enhanced or depressed on these scales. 

We also consider the variation of these statistics with changing cosmological parameters. To explore this, we show as examples in Figs. \ref{fig:gamma_squared_ratios_real_cosmologies} and \ref{fig:gamma_gammastar_ratios_real_cosmologies} the ratio with different cosmologies for the $\gamma^2$ and $\gamma^*\gamma$ statistics as a function of multipole $\ell$ as compared to the fiducial Planck cosmology. We vary in particular $\sigma_8$, $\Omega_m$, $S_8 \equiv \sigma_8 \sqrt{\Omega_m/0.3}$, and $n_s$, since these are the parameters most relevant for weak lensing studies. We see that in Fig.~\ref{fig:gamma_gammastar_ratios_real_cosmologies} the feature around $\ell\sim 370$ comes from the division of the statistics when the quantities are near to zero (see Fig.~\ref{fig:angular_power}). We highlight the difference between the $\gamma^2$ and $\gamma^*\gamma$ statistics, where we note that $\gamma^*\gamma$ shows a larger difference, particularly for larger $\ell$, when compared to $\gamma^2$, for each of the variations in parameters.  We note that there is a simple dependence on cosmology, making their use in inference pipelines more feasible since we should be able to interpolate between cosmologies to produce a dense parameter space necessary for reliable inference.

\section{Discussion}\label{sec:discussion}
We have presented a generalization of skew-spectra to include arbitrary compositions of spin-$s$ fields, and applied this formalism to weak lensing, where we presented and analyzed the relevant statistics in $\Lambda$CDM. We found that the skew-spectra in general involve a complex linear combination of types of bispectra ($EEE$, $EEB$, ...). We discussed the case where we combine arbitrary fields and filters and explained that these types of bispectra in general are not able to be isolated. This is possible, however, when considering statistics formed of $a\times a^2$, where $a$ is an arbitrary spin-$s$ field, or if we filter the fields such that only one $\ell$-space triangle contributes to the statistics. The validation of these techniques in the presence of complex survey geometries will be the subject of future work. 

In the context of weak lensing studies, we showed that the relevant statistics are formed by constructing the $\gamma^2$, $\gamma^*\gamma$, $\delta_g\gamma$ quantities. We emphasized that this has the advantage of not involving mass-mapping techniques which are often used in other higher-order weak lensing statistics. We then made predictions for the statistics using the BiHaloFit 
fitting function and examined the form of the statistics as a function of scale. The overall shape and amplitude of these statistics are largely driven by the behavior of the Wigner-$3j$ symbols, which account for features such as suppression or enhancement of power at scales near the filter edges. Finally, we discussed the cosmology dependence of the statistics, showing examples for the $E^{\gamma}\times E^{\gamma^2}$ and $E^{\gamma}\times E^{\gamma^*\gamma}$ statistics for variations in $\sigma_8$, $\Omega_m$, $S_8$ and $n_s$. 

We plan to apply this formalism to weak-lensing data to assess the viability of incorporating these statistics into cosmological analysis pipelines. In particular, it remains to be determined which statistics achieve sufficiently high signal-to-noise, remain robust in the presence of complex survey geometries, have a simple covariance structure, and provide significantly increased constraining power relative to traditional `$3$x$2$ point' (galaxy-galaxy, galaxy-lensing and lensing-lensing power spectra) analyses. After these questions are addressed, we anticipate that these statistics can be incorporated smoothly into cosmological analysis pipelines, benefiting from the extensive existing infrastructure for angular power spectrum calculations. Moreover, we point out that the generality of the formalism suggests that it may be useful for CMB polarization analyses.

\section*{Acknowledgements}
We thank Léa Harscouët for useful comments. AR acknowledges funding from STFC. SM acknowledges support from the Beecroft Trust. DA acknowledges support from STFC and the Beecroft Trust. PGF acknowledges support from STFC and the Beecroft Trust.

\appendix
\section{Spin-weighted spherical harmonics\label{sec:spin_fields}}
In the same way that spherical harmonics form a complete basis for functions defined on the sphere, the spin-$s$ weighted spherical harmonics form a complete basis for a spin-$s$ fields on the sphere. We define the SWSHs as 
\begin{equation}
    _{s}Y_{lm}\equiv \sqrt{\frac{(l-s)!}{(l+s)!}}\eth^{s}Y_{lm},
\end{equation}
\begin{equation}
    _{-s}Y_{lm}\equiv (-1)^{s}\sqrt{\frac{(l-s)!}{(l+s)!}}\bar{\eth}^{s}Y_{lm},
\end{equation}
where $\eth$ and $\bar{\eth}$ are defined in Eqs.~(\ref{eq:spin_raising}) and (\ref{eq:spin_lowering}) respectively. They are zero for $|s|> l $ or $|m|>l$. We list various useful results for the SWSHs
\begin{equation}\label{eq:spherical_harmonics_id}
    _{s}Y^{*}_{lm}=(-1)^{m+s} {}_{-s}Y_{l-m},
\end{equation}
\begin{equation}
    \eth_{s}Y_{lm} = \sqrt{(l-s)(l+s+1)} {}_{s+1} Y_{lm},
\end{equation}
\begin{equation}
    \bar{\eth}_{s}Y_{lm} = \sqrt{(l+s)(l-s+1)} {}_{s-1} Y_{lm},
\end{equation}
\begin{equation}\label{eq:spin_harmonics_five}
    \bar{\eth}^{p} \eth^{p} {}_{s}Y_{lm} = (-1)^{p}\frac{(l-s)!(l+s+p)!}{(l-s-p)!(l+s)!} {}_{s}Y_{lm},
\end{equation}
\begin{equation}\label{eq:orthogonality_SWSH}
    \int \mathrm{d}\nv {}_{s}Y_{lm}(\nv) {}_{s}Y^{*}_{l'm'}(\nv) = \delta_{ll'}\delta_{m m'
    },
\end{equation}
\begin{equation}
    \mathbb{P}(_{s}Y_{\ell m}) = (-1)^{\ell} \ {}_{-s}Y_{\ell m}.
\end{equation}
We can find a coupling relation between the SWSH 
\begin{widetext}
\begin{equation}\label{eq:spin_coupling}
\begin{split}
    &{}_{s_1} Y_{\ell_1 m_1}(\nv)\ {}_{s_2} Y_{\ell_2 m_2}(\nv) = \sqrt{\frac{(2\ell_1+1)(2\ell_2+1)(2\ell+1)}{4\pi}} \sum_{\ell, m  } (-1)^{m+s_1+s_2}\begin{pmatrix}
        \ell_1 & \ell_2 & \ell \\
        m_1 & m_2 & -m
    \end{pmatrix}
    \begin{pmatrix}
        \ell_1 & \ell_2 & \ell \\
        -s_1 & -s_2 & s_1+s_2
    \end{pmatrix} {}_{s_1+s_2}Y_{\ell m}(\nv), 
\end{split}
\end{equation}
and using the orthogonality of the SWSHs in Eq.~(\ref{eq:orthogonality_SWSH}) we can re-write this in a more familiar form
\begin{equation}\label{eq:integral_triple_swsh}
\begin{split}
    &\int \mathrm{d}\nv \ {}_{s_1}Y_{\ell_1 m_1}(\nv) \ {}_{s_2}Y_{\ell_2 m_2}(\nv) \ {}_{s_3}Y_{\ell_3 m_3}(\nv) = \sqrt{\frac{(2\ell_1+1)(2\ell_2+1)(2\ell_3+1)}{4\pi} } \begin{pmatrix}
        \ell_1 & \ell_2 & \ell_3 \\
        s_1 & s_2 & s_3
    \end{pmatrix}
    \begin{pmatrix}
        \ell_1 & \ell_2 & \ell_3 \\
        m_1 & m_2 & m_3
    \end{pmatrix},
\end{split}
\end{equation}
\end{widetext}
where the matrices represent the Wigner-$3j$ symbols. This reduces to the standard Gaunt integral when setting $s_1=s_2=s_3=0$. The Clebsch-Gordan coefficients are related to the Wigner-$3j$ symbols as follows
\begin{equation}\label{eq:clesch_wigner}
    \langle \ell_1 \ell_2 m_1 m_2| \ell_1 \ell_2 \ell m \rangle = (-1)^{\ell_2-\ell_1-m}\sqrt{2\ell+1} \begin{pmatrix}
        \ell_1 & \ell_2 & \ell \\
        m_1 & m_2 & -m
    \end{pmatrix}.
\end{equation}The Wigner-$3j$ symbols satisfy, among other results, the orthogonality relation
\begin{equation}\label{eq:orthogonal}
    \sum_{m_1m_2}(2\ell+1) \begin{pmatrix}
        \ell_1 & \ell_2 & \ell\\
        m_1 & m_2 & m
    \end{pmatrix} \begin{pmatrix}
        \ell_1 & \ell_2 & \ell'\\
        m_1 & m_2 & m'
    \end{pmatrix}
    = \delta_{\ell \ell'} \delta_{mm'}
\end{equation}
and the time reversal property
\begin{equation}\label{eq:time_reversal}
    \begin{pmatrix}
        \ell_1 & \ell_2 & \ell\\
        -m_1 & -m_2 & -m
    \end{pmatrix} = (-1)^{\ell_1+\ell_2+\ell} \begin{pmatrix}
        \ell_1 & \ell_2 & \ell\\
        m_1 & m_2 & m
    \end{pmatrix}.
\end{equation}

\section{$E/B$-squared maps}\label{sec:appendix_eb_squared}

In spite of the advantages of the approach described in the previous section, it is clear that the act of squaring the spin-$s$ field introduces a complicated sum of types of bispectra which contribute to the new statistics. We develop and explore another construction which accesses bispectrum information and avoids this complication. The idea is to `square' at the $E/B$ level, as opposed to at the spin-$s$ field level. We would make a quadratic map $F^{a}(\nv)G^{b}(\nv)$, where $F^a$ and $G^b$ denote the $E$ or $B$ modes of the spin-$s$ fields $a$ and $b$ (with spins $s_a$ and $s_b$). We then calculate the power spectra of this new map with another map as follows
\begin{equation}\label{eq:skew_spectrum_eb_squared}
    \langle F^{a}_{\ell' m'} (G^{b}H^{c})_{\ell m}\rangle  \equiv (-1)^{m}\delta_{\ell\ell'}\delta_{m-m'} C_{\ell}^{F^{a}  (G^{b} H^{c})}.
\end{equation}
The quadratic maps can be expanded as
\begin{equation}
    \begin{split}
    &F^{a}(\nv)G^{b}(\nv) =  \sum_{\ell_1 m_1 \ell_2m_2}  F^{a}_{\ell_1 m_1}G^{b}_{\ell_2 m_2} Y_{\ell_1 m_1 }(\nv)Y_{\ell_2 m_2 }(\nv)\\
        &=\sum_{\ell_1 m_1\ell_2 m_2}\sum_{\ell m}F^{a}_{\ell_1 m_1} G^{b}_{\ell_2 m_2}(-1)^mY_{\ell m}(\nv)\begin{pmatrix}
        \ell_1 & \ell_2 & \ell\\
        0 & 0 & 0
        \end{pmatrix}\times \\
        &\times\begin{pmatrix}
           \ell_1 & \ell_2 & \ell\\
           m_1  & m_2 & -m
        \end{pmatrix} \sqrt{\frac{(2\ell_1+1)(2\ell_2+1)(2\ell+1)}{4\pi }}.
    \end{split}
\end{equation} 
We can identify the harmonic coefficients of these maps. We proceed to calculate its relation to the angular bispectra as we did for the spin-squared maps. We then simply take the product of this with the $E/B$ modes, use the definition of the angular bispectra, Eq.~(\ref{eq:full_bispectrum}), and the orthogonality relation of the Wigner $3j$ symbols, Eq.~(\ref{eq:orthogonal}), to calculate the skew-spectra of the $E/B$-squared maps (Eq.~(\ref{eq:skew_spectrum_eb_squared}))
\begin{equation}
     C_{\ell}^{F^a(G^bH^c)} = \sum_{\ell_1\ell_2 } \sqrt{\frac{(2\ell_1+1)(2\ell_2+1)}{4\pi(2\ell+1)}}\begin{pmatrix}
  \ell_1 & \ell_2 & \ell\\
    0 & 0 & 0 
\end{pmatrix}b^{F^aG^bH^c}_{\ell \ell_1 \ell_2 }.
\end{equation}
 We note the similar mathematical structure to the scalar field skew-spectrum of Ref.~\cite{Harscouet2025Fast}, due to its nature as combining scalar quantities. We see that only one type of bispectrum contributes to the building of the new statistics. This makes its interpretation much clearer than the spin-squared counterpart. However, we also see that the Wigner-$3j$ means that it only receives contributions from $\ell_1+\ell_2+\ell=\text{even}$ bispectrum configurations. The key significant disadvantage of this is that the $E$ and $B$ fields are non-local, leading to leakage effects which are amplified by squaring. It is not clear how these quantities can be corrected for these mask effects. \\

\newpage

\bibliography{main}

\end{document}